\documentclass[aps,prl,twocolumn,superscriptaddress]{revtex4-1}
\usepackage{epsfig,pstricks,graphicx}
\usepackage{amsmath,amssymb}
\usepackage[mathscr]{eucal}
\usepackage{color}
\usepackage[normalem]{ulem}

\def\CC{{\rm\kern.24em \vrule width.04em height1.46ex depth-.07ex
\kern-.30em C}}
\def\RR{{\rm
         \vrule width.04em height1.58ex depth-.0ex
         \kern-.04em R}}
\def\P{{\rm I\kern-.25em P}}
\def\id{{\rm 1\kern-.22em l}}

\newcommand{\bra}[1]{\left\langle #1 \right |}
\newcommand{\ket}[1]{\left | #1 \right\rangle}

\newcommand{\oost}{\frac{1}{\sqrt{2}}}
\newcommand{\rme}{\ensuremath{\mathrm{e}}}
\newcommand{\rmi}{\ensuremath{\mathrm{i}}}
\newcommand{\rmd}{\ensuremath{\mathrm{d}}}

\newcommand{\tr}{\operatorname{tr}}
\newcommand{\re}{\operatorname{Re}}

\begin{document}

\title{ A quantitative witness for Greenberger-Horne-Zeilinger entanglement
      }
\author{Christopher Eltschka}
\affiliation{Institut f\"ur Theoretische Physik, 
         Universit\"at Regensburg, D-93040 Regensburg, Germany}
\author{Jens Siewert}
\affiliation{Departamento de Qu\'{\i}mica F\'{\i}sica, Universidad del Pa\'{\i}s Vasco UPV/EHU,
             E-48080 Bilbao, Spain}
\affiliation{IKERBASQUE, Basque Foundation for Science, E-48011 Bilbao, Spain}
\begin{abstract}
Along with the vast progress in experimental quantum technologies 
there is an increasing demand for the quantification of entanglement 
between three or more quantum systems. Theory still does not provide 
adequate tools for this purpose. 
The objective 
is, besides the quest for exact results, to develop operational 
methods that allow for efficient entanglement quantification. 
Here we put forward an analytical approach that serves both these goals. 
We provide a simple procedure to
quantify Greenberger-Horne-Zeilinger--type multipartite entanglement
in arbitrary three-qubit states. 
For two qubits this method is equivalent to Wootters' seminal 
result [Phys.\ Rev.\ Lett.\ {\bf 80}, 2245 (1998)].
It establishes a close link between entanglement quantification 
and entanglement detection by witnesses,
and can be generalised both to higher dimensions and to more 
than three parties.
\end{abstract}

\maketitle
%

It is a fundamental strength of physics as a science that most
of its basic concepts have quantifiability built into their definition. 
Just think of, {\em e.g.}, length, time, or electrical current.
Their quantifiability allows to measure and compare 
them in different contexts, and to build mathematical theories with them~\cite{Helmholtz1887}.
There is no doubt that entanglement is a key concept in quantum theory,
but it seems to resist in a wondrous way that universal principle of quantification.
The reason for this is, in the first place, that entanglement comes in many
different disguises related to its resource character, {\em i.e.}, 
what one would like to do with it. 
In principle, 
there are numerous task-specific entanglement measures~\cite{Plenio2007,review2}. 
However, most of them cannot be calculated easily 
(nor measured or estimated) for generic mixed quantum states, 
and therefore it is difficult to use them. 

There are notable exceptions, the concurrence~\cite{Bennett1996} and the negativity 
for bipartite systems~\cite{Vidal2002}. These measures have already provided 
deep insight into the nature of entanglement, but they also have their shortcomings. 
The concurrence is strictly applicable only to two-qubit systems while for 
the negativities it is not known how to distinguish
entanglement classes.
The generalisations of the concurrence (such as the residual tangle~\cite{CKW2000})
do quantify task-specific entanglement even 
for multipartite systems but again it is not known how to estimate them for
general mixed quantum states.

There is another difficulty. 
An $N$-qubit density matrix $\rho$ is characterised mathematically by $2^{2N}-1$ real 
parameters. Reducing it to its so-called 
normal form~\cite{Verstraete2003}---which contains the essential 
entanglement information---removes about $6N$ parameters. 
The entanglement measure is determined 
by the remaining exponentially many parameters which need to be processed 
to calculate the precise value. Even an operational method similar to that of
Wootters-Uhlmann~\cite{Wootters1998,Uhlmann2000} 
would quickly reach its limits
with increasing $N$.  Therefore it is desirable to develop methods
which provide useful approximate answers even for larger systems.
If one asks for mere entanglement detection, witnesses~\cite{GuehneToth2009} 
are such a tool because
here the number of required parameters (both for measurement and processing)
can be reduced substantially. 
There are also estimates of  entanglement measures using witness 
operators~\cite{Reimpell2007,Eisert2007} which,
however,  have not yet produced practical methods for entanglement
quantification.

Here we develop an easy-to-handle quantitative witness 
for Greenberger-Horne-Zeilinger (GHZ) entanglement~\cite{Duer2000}
in arbitrary three-qubit states.
It yields the exact three-tangle for the family of 
GHZ-symmetric states~\cite{Eltschka2012}, and those states which are locally 
equivalent to them.  For all other states, the method gives an optimised
lower bound to the three-tangle. Due to this feature we call the approach 
a {\em witness}.

We start by defining the GHZ symmetry~\cite{Eltschka2012} 
and stating our central result.
Then we prove the validity of the 
statement for two qubits. We obtain a method 
equivalent to that of Wootters-Uhlmann, {\em i.e.}, it gives the
exact concurrence for arbitrary density matrices. Subsequently we
explain the extension of the approach to arbitrary three-qubit states.

\section{Results}

\subsection{The procedure}

The $N$-qubit GHZ state in the computational basis is defined as
$\ket{\mathrm{GHZ}}\equiv\frac{1}{\sqrt{2}}(\ket{00\ldots 0}+\ket{11\ldots 1})$.
It is invariant under: {\em (i)} Qubit permutation.
{\em (ii)} Simultaneous spin flips {\em i.e.}, application of 
$\sigma_x^{\otimes N}$.
{\em (iii)}~Correlated local $z$ rotations:
\begin{equation}
    \label{eq:zrotNqb}
  U_N = \rme^{\rmi \varphi_1 \sigma_z}\otimes
                              \rme^{\rmi \varphi_2 \sigma_z}\otimes\ldots
                             \rme^{-\rmi \left(\sum_1^{N-1}\varphi_j\right) \sigma_z}
  \end{equation}
where $\sigma_x$, $\sigma_y$, $\sigma_z$ are Pauli matrices.
An $N$-qubit state is called
{\em GHZ symmetric} and denoted by $\rho^{\mathrm{S}}$
if it remains invariant under the operations {\em (i)-(iii)}.
An arbitrary $N$-qubit state $\rho$ can be symmetrized by the operation
\begin{equation}
\label{eq:symmstate}
 \rho^{\mathrm{S}}(\rho) = \int\rmd U_{\mathrm{GHZ}}\, U_{\mathrm{GHZ}} 
                           \rho\, U_{\mathrm{GHZ}}^\dagger
\end{equation}
where the integral 
denotes averaging over 
the GHZ symmetry group including permutations and
spin flips. Notably, the GHZ-symmetric $N$-qubit states form a convex 
subset of the space of all $N$-qubit states.
\\
{\bf Observation:} 
{\em 
If an appropriate entanglement measure $\mu$ is known exactly for 
GHZ-symmetric N-qubit states $\rho^{\mathrm{S}}$, it can be employed to
quantify GHZ-type entanglement in arbitrary 
$N$-qubit states $\rho$.
Here, 
$\mu(\psi)$ is a positive $\mathrm{SL}(2,\CC)^{\otimes N}$-invariant function
of homogeneous degree 2 in the coefficients of a pure quantum state $\psi$,
and $\mu(\rho)$ is its convex-roof extension~\cite{Uhlmann1998}.
The estimate for $\mu(\rho)$ is found in the following sequence of steps:\\
(1) Given  a state $\rho$, derive a normal form 
    $\rho^{\mathrm{NF}}(\rho)$.
    If $\rho^{\mathrm{NF}}(\rho)=0$ the procedure terminates here,
    and $\mu(\rho)=0$.\\
(2) Renormalise $\rho^{\mathrm{NF}}/\tr{\rho^{\mathrm{NF}}}$
    and transform it 
    using local unitaries $\mathrm{SU}(2)^{\otimes N}$
    to the state $\tilde{\rho}^{\mathrm{NF}}(\rho)$ 
     according to appropriate criteria (see below)
     so that the entanglement of $\rho^{\mathrm{S}}(
          \rho^{\mathrm{NF}}/\tr{\rho^{\mathrm{NF}}})$ is enhanced.
    \\
(3) Project the state onto the GHZ-symmetric states
    $\tilde{\rho}^{\mathrm{NF}}(\rho) \mapsto
    \rho^{\mathrm{S}}(\tilde{\rho}^{\mathrm{NF}})$. 
    The estimate for $\mu(\rho)$ is obtained after renormalisation
\[
    \mu(\rho^{\mathrm{S}}(\tilde{\rho}^{\mathrm{NF}}))\tr\rho^{\mathrm{NF}}\!
    \leq \mu(\rho)  \ \ .
\]
}

\subsection{Two qubits}

For two qubits the entanglement measure under consideration is the 
concurrence $C(\rho)$ (Refs.~\cite{Bennett1996,Wootters1998}).
From the symmetrization $\rho^{\mathrm{S}}(\rho)$ of an arbitrary 
two-qubit state $\rho$ we find~(for details see Supplementary Information):
\begin{equation}
     C(\rho) \geq \max{\left( 0 , |\rho_{00,11}+\rho_{11,00}| + \rho_{00,00}+\rho_{11,11} - 1\right)}\ \ .
\label{eq:2qbconcrho}
\end{equation}
In the symmetrization entanglement may be lost, as
illustrated by the state
$\ket{\Psi^-}=\oost (\ket{01}-\ket{10})$  for which 
inequality~\eqref{eq:2qbconcrho}
gives the poor estimate $C(\Psi^-)\geq 0$.
Therefore, the optimisation steps {\em (1)} and {\em (2)} 
are necessary to avoid unwanted entanglement loss in
the symmetrization {\em (3)}.  The goal is
to augment the right-hand side of inequality~\eqref{eq:2qbconcrho} up to the
point that equality is reached. We will show now that for two qubits this can 
indeed be achieved.

It is fundamental that the maximum of an SL$(2,\CC)^{\otimes N}$-invariant 
function $\mu(\rho)$ under general local operations can be reached by 
applying the optimal transformation
$\rho \mapsto A\rho A^{\dagger}/
       \tr A\rho A^{\dagger}$
where $A=A_1\otimes\ldots\otimes A_N$ and $A_j\in  \mathrm{SL}(2,\CC)$
is an invertible local operation~\cite{Verstraete2003}.
Consider first the normal form $\rho^{\mathrm{NF}}(\rho)$ which 
is obtained from $\rho$
by iterating determinant-one local operations~\cite{Verstraete2003}
 (see also Methods).  
Such operations (represented by SL$(2,\CC)$ matrices) describe
stochastic local operations and classical communication
(SLOCC). Consequently, the normal form is locally equivalent to the
original state $\rho$, that is, it lies in the entanglement class of $\rho$. 
Note that the iteration leading to the normal form {\em minimises} 
the trace of the state.  Subsequent renormalisation 
{\em increases} the absolute values of all matrix elements in 
equation~\eqref{eq:2qbconcrho}.
Here, the correct rescaling of the mixed-state entanglement measure is crucial. 
This is why homogeneity degree 2 of $\mu(\psi)$ is required~\cite{Gour2010,Viehmann2012APB}.

Hence, transforming $\rho$ to its normal form increases the moduli of 
$\rho_{00,00}$, $\rho_{00,11}$,  $\rho_{11,00}$,  $\rho_{11,11}$ (and also the
concurrence) as much as possible
for a state that is SLOCC equivalent with $\rho$. 
The sum of the off-diagonal matrix elements
in  equation~\eqref{eq:2qbconcrho} reaches its maximum
if $\rho_{00,11}$ is real and positive. 
As this can be achieved by a $z$ rotation on
one qubit we may consider it  part of finding the normal form and
drop the absolute value bars in equation~\eqref{eq:2qbconcrho}. 
Then, the sum of matrix elements 
equals, up to a factor 1/2, the fidelity of 
$\rho^{\mathrm{NF}}/\tr{\rho^{\mathrm{NF}}}$ with the Bell state 
$\ket{\Phi^+}=\frac{1}{\sqrt{2}}(\ket{00}+\ket{11})$. The question is how large
this fidelity may become.

To find the answer 
we transform $\rho^{\mathrm{NF}}/\tr\rho^{\mathrm{NF}}$ 
to a Bell-diagonal form using local unitaries 
(this is always possible~\cite{Verstraete2001,Verstraete2003,Leinaas2006}).
If then  $\rho^{\mathrm{NF}}_{00,11}<|\rho^{\mathrm{NF}}_{01,10}|$ 
we apply another SU(2)$^{\otimes 2}$ operation 
to maximise $\rho^{\mathrm{NF}}_{00,11}$ (see Supplementary Information). 
The result is
a Bell-diagonal $\tilde{\rho}^{\mathrm{NF}}$ with maximum real off-diagonal element 
$\tilde{\rho}^{\mathrm{NF}}_{00,11}$. However,  Bell-diagonal two-qubit
density matrices with this property can be made
GHZ symmetric {\em without} losing
entanglement~\cite{Bennett1996} (see also Supplementary Information).

Hence, our optimised symmetrization
procedure {\em (1)-(3)} leads to the exact concurrence for
arbitrary two-qubit states $\rho$. 
In passing, we have demonstrated that the concurrence is related 
via $C(\rho)=\max(0,2f-1)\cdot\tr\rho^{\mathrm{NF}}$ to the maximum fidelity 
	$f= \bra{\Phi^+}\tilde{\rho}^{\mathrm{NF}} \ket{\Phi^+}$
that can be achieved by
applying invertible local operations to $\rho$.

\subsection{Three qubits}

For three qubits, the GHZ-symmetric states are described by two 
parameters~\cite{Eltschka2012} and therefore form a 
two-dimensional submanifold in the space of all three-qubit 
density matrices. It turns out that it has the shape of a flat
isosceles triangle, see Fig.~1. A convenient parametrisation is 
\begin{eqnarray}
\label{eq:coords3qbx}
    x(\rho)\ &=&\ \frac{1}{2} \left( \rho_{000,111}+\rho_{111,000}  \right)
\\
\label{eq:coords3qby}
    y(\rho)\ &=&\ \frac{1}{\sqrt{3}}\left(\rho_{000,000}+\rho_{111,111}
                                                     -\frac{1}{4}\right)
\end{eqnarray}
as it makes the Hilbert-Schmidt metric in the space of density
matrices conincide with the Euclidean metric. This way geometrical
intuition can be applied to understand the properties of this set
of states. All entanglement-related properties of GHZ-symmetric states
are symmetric under  sign change $x \leftrightarrow -x$ as this is
achieved by applying $\sigma_z$ to one of the qubits.
%
\begin{figure}[bht]
  \centering
  \includegraphics[width=.95\linewidth]{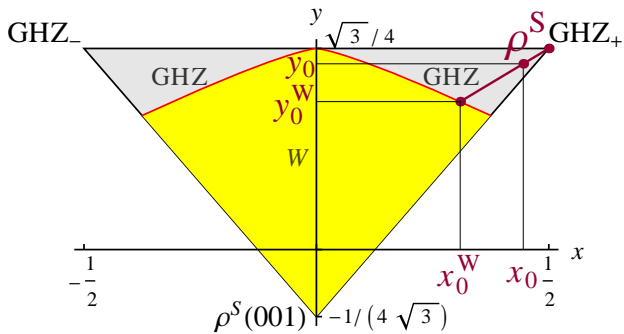}
  \caption{The triangle of GHZ-symmetric three-qubit states.
           The upper corners  correspond to 
           $\ket{\mathrm{GHZ}_{\pm}}\equiv\frac{1}{\sqrt{2}}(\ket{000}\pm\ket{111})$  
           and the lower corner to $\rho^{\mathrm{S}}(001)$, cf.\ Ref.~\cite{Eltschka2012}. 
           The grey area shows GHZ-class states ($\tau_3>0$) whereas
           the yellow area comprises states with vanishing $\tau_3$ ("$W$").
    The border between GHZ-class and $W$-class states is the
    GHZ/$W$ line, equation~\eqref{SIeq:b-W} (red solid line). 
           We also show a state $\rho^{\mathrm{S}}(x_0,y_0)$
           together with the point $(x_0^W,y_0^W)$ that is required to determine
           the three-tangle $\tau_3(x_0,y_0)$, 
           equation~\eqref{eq:result3}. 
    }
\end{figure}
%

The GHZ-class entanglement of  three-qubit states is 
quantified by the  three-tangle  $\tau_3$ (Refs.~\cite{CKW2000,Viehmann2012APB}, 
see also Methods). 
For GHZ-symmetric three-qubit states $\rho^{\mathrm{S}}(x_0,y_0)$
the exact solution for the three-tangle~\cite{Siewert2012} (see also
Methods) is 
\begin{equation}
   \tau_3(x_0,y_0) =
  \begin{cases} \ \ \
    0 \ \ \ \ \ \ \ \ \ \ \mbox{for } x_0<x^W_0\ \mbox{and }y_0<y^W_0\\[2mm]
   \displaystyle
   \frac{x_0-x^W_0}{\frac{1}{2}-x_0^W}=
   \frac{y_0-y^W_0}{\frac{\sqrt{3}}{4}-y^W_0}  
                  \ \ \ \ \ \           \mbox{otherwise}
  \end{cases}
   \label{eq:result3}
\end{equation}
where $x_0\geq 0$
and $(x_0^W,y_0^W)$ are the coordinates of the intersection of
the GHZ/$W$ line with the direction that contains both GHZ$_+$ 
and $\rho^{\mathrm{S}}(x_0,y_0)$ (cf.~Fig.~1). The grey surfaces
in Fig.\ 2 illustrate this solution.

Now we turn to constructing a quantitative witness for the three-tangle
of arbitrary three-qubit states by using the solution in 
equation~\eqref{eq:result3}. As before, the main idea is that an arbitrary
state can be symmetrized according to equation~\eqref{eq:symmstate}
and thus is projected into the GHZ-symmetric states.
Again, we assume $\rho_{000,111}$ real and nonnegative, so that $x(\rho)\geq 0$.
From Figs.~1~and~2 it appears evident that the entanglement of the symmetrization image
$\rho^{\mathrm{S}}(\rho)$ can be improved by moving its point $(x(\rho),y(\rho))$
closer to GHZ$_+$. 
More precisely, the entanglement measure is enhanced
upon increasing one of the coordinates without decreasing the other 
(cf.~equations~\eqref{eq:2qbconcrho} and \eqref{eq:result3}).   

In this spirit, finding the normal form in step {\em (1)} is appropriate
as it yields the largest possible three-tangle for a state 
$\rho^{\mathrm{NF}}/\tr\rho^{\mathrm{NF}}$
locally equivalent to the original $\rho$ (cf.~Ref.~\cite{Verstraete2003}). 
As the normal form is unique only up to local unitaries it does not
automatically give the state with minimum entanglement loss in the 
symmetrization.
Therefore, the unitary optimisation step {\em (2)} is required to generate 
the best coordinates.

In the symmetrization the information contained in various  matrix
elements is lost.
For two qubits, however, the concurrence of the optimised Bell-diagonal
states depends only on 
$\tilde{\rho}_{00,00}^{\mathrm{NF}}+\tilde{\rho}_{00,11}^{\mathrm{NF}}$
and the loss of $\tilde{\rho}_{01,10}^{\mathrm{NF}}$ in the symmetrization does not harm. 
In contrast, the three-qubit normal form
depends on about 
45 parameters.
We may not expect that $\tau_3(\rho)$ depends only on two of them and,
hence, entanglement loss in the symmetrization {\em (3)}
is inevitable~(cf.~Supplementary Information). 
Consequently, steps {\em (1)-(3)}
lead to a lower bound for the three-tangle
that coincides with the exact $\tau_3(\rho)$  at least for those states which
are locally equivalent to a GHZ-symmetric state. 
The most straightforward optimisation criterion in step {\em (2)}
is to maximise $\mu(\rho^{\mathrm{S}}(\tilde{\rho}^{\mathrm{NF}}))$.
Alternative criteria which generally do not give the best
$\tau_3(\rho)$ but can be handled more easily (possibly analytically) 
are maximum fidelity
$\bra{\mathrm{GHZ}_+}\rho^{\mathrm{S}}
(\tilde{\rho}^{\mathrm{NF}})\ket{\mathrm{GHZ}_+}$,  minimum
Hilbert-Schmidt distance of 
$\rho^{\mathrm{S}} (\tilde{\rho}^{\mathrm{NF}})$ from GHZ$_+$, 
or  maximum $\re\tilde{\rho}_{0\ldots 0,1\ldots 1}^{\mathrm{NF}}$.
%
%
%
\begin{figure}[htb]
  \centering
  \includegraphics[width=.95\linewidth]{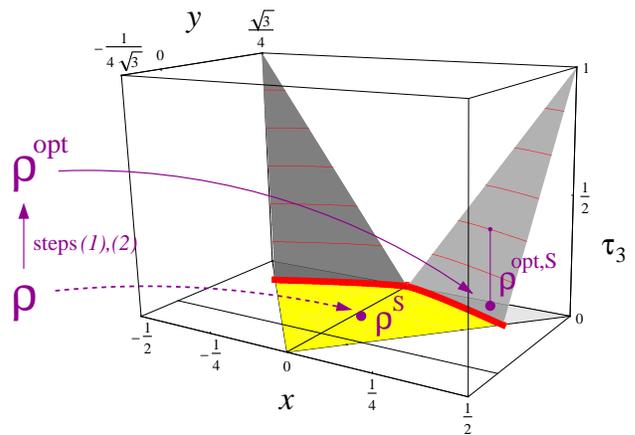}
  \caption{
           Illustration of the procedure for finding the three-tangle
           of a general mixed three-qubit state $\rho$.
           In the $xy$ plane, there is the triangle of GHZ-symmetric states
           while on the vertical axis, the three-tangle for 
           each GHZ-symmetric state (cf.~equation~\eqref{eq:result3})) is shown.
           Simple projection $\rho \mapsto \rho^{\mathrm{S}}$ generates
           a non-optimal GHZ-symmetric state. The optimisation steps 
           {\em (1), (2)} move the symmetrization image to  
           $\rho^{\mathrm{opt,S}}\equiv\rho^{\mathrm{S}}(\tilde{\rho}^{\mathrm{NF}})$ 
           with enhanced three-tangle.
            } 
\end{figure}
%

\section{Discussion}

Evidently
this approach can be generalised. Therefore we conclude
with a discussion of some of its universal features.
The essential ingredients are an exact solution of
the entanglement measure for a sufficiently general family of states 
with suitable symmetry, and the entanglement optimisation
for a given arbitrary state $\rho$ via general local operations. 
The former determines the border where the
entanglement vanishes. The latter ensures an appropriate fidelity 
of the image $\rho^{\mathrm{S}}(\rho)$ with the maximally entangled state.
This reveals a remarkable relation between entanglement quantification 
through SL$(2,\CC)$ invariants and the standard entanglement witnesses
which we briefly explain in the following.

A well-known witness for two-qubit entanglement is
$\mathcal{W}_2=\frac{1}{2}\id_4-\ket{\Phi^+}\!\bra{\Phi^+}$. 
It detects the entanglement of an arbitrary normalised two-qubit state 
$\rho^{\mathrm{2qb}}$ if
\[
       0 \ >\   \tr\left(\rho^{\mathrm{2qb}}\mathcal{W}_2 
               \right)\ =\  \frac{1}{2} - 
                            \bra{\Phi^+} \rho^{\mathrm{2qb}}\ket{\Phi^+}
\ \ .
\]
On the other hand, 
from our concurrence result 
\begin{align}
       C(\rho^{\mathrm{2qb}}) & =  \max \left(0,\max_{A=A_1\otimes A_2}
         \left[2\bra{\Phi^+} A \rho^{\mathrm{2qb}}
                              A^{\dagger}\ket{\Phi^+} - 
\right.\right.   \nonumber\\
      &   \left.\left. \mbox{ }\hspace*{3.9cm} 
            - \tr \left(A \rho^{\mathrm{2qb}}
                        A^{\dagger}\right)
                                                                \right]
                                       \right)
\nonumber\\
      & \ge  \ 2\bra{\Phi^+}  \rho^{\mathrm{2qb}} \ket{\Phi^+}
                 - \tr \left(\rho^{\mathrm{2qb}}
                       \right)
\\
      & = \ -2 \tr\left(\rho^{\mathrm{2qb}}\mathcal{W}_2\right)
\nonumber
\end{align}
we see, by dropping the optimisation over SLOCC 
$A=A_1\otimes A_2$,
 that $\mathcal{W}_2^{\prime}=-2\mathcal{W}_2$ is a (non-optimised) 
quantitative witness 
for two-qubit entanglement.  In other words, $\mathcal{W}_2^{\prime}$ yields
one of the many possible lower bounds to the exact result.  
Analogously it is straightforward to establish the relation between the 
standard GHZ witness 
$\mathcal{W}_3=\frac{3}{4}\id_8-\ket{\mathrm{GHZ}_+}\!\bra{\mathrm{GHZ}_+}$ 
and the non-optimal quantitative witness 
$\mathcal{W}_3^{\prime}=-4\mathcal{W}_3$. The latter represents a
linear lower bound to the three-tangle obtained via the optimisation
steps {\em (1)--(3)}~(see Supplementary Information).

Finally we mention that our approach can be used without optimisation,
{\em i.e.}, either without step {\em (1)}, or {\em (2)}, or both.
This renders the witness less reliable but more efficient.
At best it requires  only four matrix elements (for {\em any} $N$).
We note  that, if we apply the witness to a tomography outcome 
the measurement effort can be reduced by using the prior knowledge of the state and 
choosing the local measurement directions such that the fidelity with the
expected GHZ state is measured directly. This implements optimisation step 
{\em (2)} right in the measurement.

\section{Methods}

\subsection{Normal form of an $N$-qubit state}

The normal form of a multipartite quantum state is a fundamental concept 
that was introduced by Verstraete {\em et al.}~\cite{Verstraete2003}
It applies to arbitrary (finite-dimensional) multi-qudit states. Here we focus
on $N$-qubit states only.

In the normal form of an $N$-qubit state $\rho$, all local density matrices 
are proportional to the 
identity. Therefore the normal form is unique up to local unitaries.
Remarkably, the normal form can be obtained by applying an appropriate
{\em local filtering operation}
\[
     \rho^{\mathrm{NF}}\ =\ \textstyle{(A_1\otimes\ldots\otimes A_N)}\rho
                            \textstyle{(A_1\otimes\ldots\otimes A_N)^{\dagger}} 
\]
where $A_j \in \mathrm{SL}(2,\CC)$.
Therefore $\rho^{\mathrm{NF}}$ is locally equivalent to the original state $\rho$.
The normal form $\rho^{\mathrm{NF}}$ is peculiar since it has the {\em minimal} 
norm of all states in the orbit of $\rho$ generated by local filtering operations.
Practically, the normal form can be found by a simple iteration procedure
described in Ref.~\cite{Verstraete2003} It is worth noticing that GHZ-symmetric
states -- which play a central role in our discussion -- are naturally given
in their normal form.

\subsection{Three-tangle of three-qubit GHZ-symmetric states}
%
The pure-state entanglement
monotone 
that needs to be considered for three-qubit states 
is the three-tangle $\tau_3(\psi)$,
{\em i.e.}, the square root of the residual tangle introduced by 
Coffman {\em et al.}~\cite{CKW2000}:
\begin{eqnarray}
  \label{SIeq:three-tangle}
  \tau_3(\psi) &=& 2\sqrt{\left|d_1 - 2 d_2 + 4 d_3\right|}, 
  \nonumber
  \\
  d_1 &=& \psi_{000}^2\psi_{111}^2 + \psi_{001}^2\psi_{110}^2 + \psi_{010}^2\psi_{101}^2
  + \psi_{011}^2\psi_{100}^2                                
  \nonumber
  \\
  d_2 &=& \psi_{000}\psi_{001}\psi_{110}\psi_{111} + \psi_{000}\psi_{010}\psi_{101}\psi_{111} +
 \nonumber
 \\ &&
  + \psi_{000}\psi_{011}\psi_{100}\psi_{111}
  + \psi_{001}\psi_{010}\psi_{101}\psi_{110} + 
 \nonumber
 \\ &&
  + \psi_{001}\psi_{011}\psi_{100}\psi_{110} +
  \psi_{010}\psi_{011}\psi_{100}\psi_{101}
 \nonumber
 \\
  d_3 &=& \psi_{000}\psi_{110}\psi_{101}\psi_{011} + \psi_{100}\psi_{010}\psi_{001}\psi_{111}
\ \ .  
\end{eqnarray}
Here $\psi_{jkl}$ with $j,k,l\in\{0,1\}$ are the components of a pure three-qubit state
in the computational basis.
The three-tangle becomes an entanglement measure also for mixed states 
$\rho=\sum_j p_j\ket{\psi_j}\!\bra{\psi_j}$ via the convex-roof 
extension~\cite{Uhlmann1998} 
\begin{equation}
      \tau_3(\rho)\ =\ \min_{\mathrm{\tiny all\ decomp.}} \sum\ p_j \ \tau_3(\psi_j) \ \ ,     
\label{eq:convex_roof}
\end{equation}
{\em i.e.}, the minimum average three-tangle
taken over all possible pure-state decompositions $\{p_j,\psi_j\}$.
In general it is difficult to carry out the minimisation procedure 
in equation~\eqref{eq:convex_roof}.
For GHZ-symmetric three-qubit states, however, the convex roof of the
three-tangle can be calculated exactly
(see equation~\eqref{eq:result3}).
This solution is shown in Fig.~2 and can be understood as follows.
The border between the $W$ and the GHZ states is the GHZ/$W$ line which has the
parametrised form~\cite{Eltschka2012}
\begin{equation}
    x^W=\frac{v^5+8v^3}{8(4-v^2)}\ \ \ ,\ \ \
    y^W=\frac{\sqrt{3}}{4}\frac{4-v^2-v^4}{4-v^2}
\label{SIeq:b-W}
\end{equation}
with $-1\leq v\leq 1$. The solution for the convex roof is obtained by
connecting each point of the GHZ/$W$ line $(x^W,y^W,\tau_3=0)$ 
with the closest of the points 
$(x_{\mathrm{GHZ}_{\pm}}=\pm\frac{1}{2},
  y_{\mathrm{GHZ}_{\pm}}=\frac{\sqrt{3}}{4},\tau_3=1)$.
That is, the three-tangle is nothing but a linear interpolation  between the points of the
border between GHZ and $W$ states, and the maximally entangled states GHZ$_{\pm}$.

%
%
%
%

%
%
%
%
%
%
\section*{Acknowledgements}
This work was funded by the German Research Foundation within 
SPP 1386 (C.E.), and by Basque Government grant IT-472-10 (J.S.).
The authors thank R.\ Fazio, P.\ Hyllus, K.F.\ Renk, and A.\ Uhlmann for 
comments, and J.\ Fabian and K.\ Richter for their support.
%
%
%
%
%
%
%
%
%
 \newpage
 \noindent
\begin{center}
           {\bf SUPPLEMENTARY INFORMATION}\\
            for \\
           {\bf  ``A quantitative witness for Greenberger-Horne-Zeilinger entanglement''}\\
            by Christopher Eltschka and Jens Siewert
\end{center}

\renewcommand{\theequation}{S\arabic{equation}}
\setcounter{equation}{0}
\setcounter{page}{1}
\renewcommand{\thefigure}{S\arabic{figure}}
\setcounter{figure}{0}
\setcounter{subsection}{0}
\subsection{Two-qubit GHZ-symmetric states}
The twirling operation equation~(2) in the main text defines a family of GHZ-symmetric 
mixed states for each qubit number $N\geq 2$. The simplest case is that of two qubits.
The symmetrization $\rho^{\mathrm{S}}(\rho)$ of an arbitrary two-qubit state $\rho$ is 
characterised by two real parameters for which we choose the following 
parametrization~\cite{Siewert2012}
\begin{eqnarray}
\label{SIeq:coords2qbx}
    x(\rho)\ &=&\ \frac{1}{2} \left( \rho_{00,11}+\rho_{11,00}  \right)
\\
\label{SIeq:coords2qby}
    y(\rho)\ &=&\ \frac{1}{\sqrt{2}}\left(\rho_{00,00}+\rho_{11,11}
                                                     -\frac{1}{2}\right)
    \ \ .
\end{eqnarray}
We emphasise that these coordinates (as well as those in 
equations~\eqref{eq:coords3qbx}, \eqref{eq:coords3qby} in the main text)
are defined for normalised density matrices.
The corresponding states form a triangle in the $xy$ plane, see Supplementary
Fig.\ S1.   
\begin{figure}[hbt]
  \centering
  \includegraphics[width=.70\linewidth]{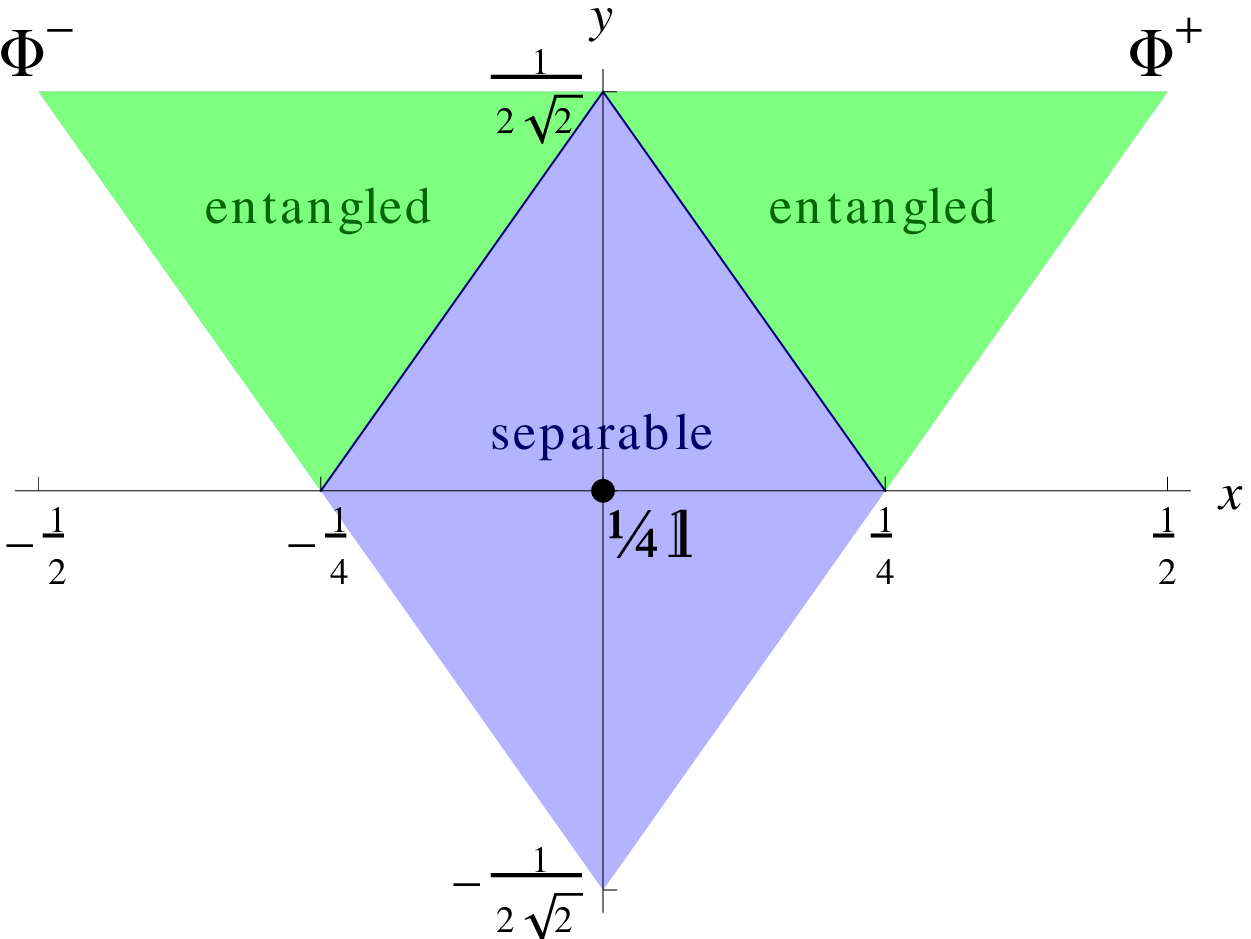}
  \caption{
    The geometric representation of two-qubit GHZ-symmetric states~\cite{Siewert2012}.
         The upper corners are defined by the Bell states
         $\ket{\Phi^-}=\frac{1}{\sqrt{2}}(\ket{00}-\ket{11})$ (left) and
         $\ket{\Phi^+}=\frac{1}{\sqrt{2}}(\ket{00}+\ket{11})$ (right).
         The lower corner represents the mixture 
         $\frac{1}{2}(\ket{\Psi^+}\!\bra{\Psi^+}+\ket{\Psi^-}\!\bra{\Psi^-})$
         with $\ket{\Psi^{\pm}}=\oost(\ket{01}\pm\ket{10})$ .
         The blue region shows
         the separable states whereas the states in the green region 
         have non-vanishing concurrence.
            } 
\end{figure}

The entanglement monotone considered here is the 
concurrence $C(\psi) = \left|\bra{\psi^{\ast}} \sigma_y\otimes\sigma_y\ket{\psi}\right|$.
Its convex-roof extension~\cite{Uhlmann1998} is defined in analogy with
equation~\eqref{eq:convex_roof} in the main text via
\begin{equation}
      C(\rho)\ =\ \min_{\mathrm{\tiny all\ decomp.}} \sum\ p_j \ C(\psi_j) \ \ ,     
\end{equation}
{\em i.e.}, the average concurrence 
minimised over all possible pure-state decompositions $\{p_j,\psi_j\}$
of the two-qubit state $\rho=\sum p_j \ket{\psi_j}\!\bra{\psi_j}$.
The concurrence of GHZ-symmetric two-qubit states
is a function of the coordinates~\cite{Siewert2012}
\begin{equation}
           C(x,y) = \max{\left( 0 , 2|x| + \sqrt{2}y - \frac{1}{2}\right)}\ \ .
\label{SIeq:2qbconcxy}
\end{equation}
We can rewrite this formula in terms of the matrix elements of the original
state $\rho$ using equations~\eqref{SIeq:coords2qbx}, \eqref{SIeq:coords2qby}, 
keeping in mind that symmetrization cannot increase the
concurrence:
\begin{equation}
     C(\rho) \geq \max{\left( 0 , |\rho_{00,11}+\rho_{11,00}| + \rho_{00,00}+\rho_{11,11} - 1\right)}
\label{SIeq:2qbconcrho}
\end{equation}
{\em i.e.}, we obtain equation~(3) of the main text.
The analogy with some of the equations in Ref.~\cite{Reimpell2007} is remarkable, 
in particular with equation (6), if we use 
$\mathcal{W}_2=\frac{1}{2}\id_4-\ket{\Phi^+}\!\bra{\Phi^+}$
as the only witness (with the optimal slope $r=-2$ and the offset $c=0$). 
 It arises due to the fact that the concurrence of 
GHZ-symmetric two-qubit states is a linear function,
and the linear one-witness
approximation in Ref.~\cite{Reimpell2007} becomes exact.
We note also that our concurrence formula in the main text, 
$C(\rho)=\max(0,2f-1)\tr\rho^{\mathrm{NF}} $,
is reminiscent of the so-called 
{\em fully entangled fraction}~\cite{Bennett1996}. However, the optimisation
of the fully entangled fraction includes only local unitaries  while
our approach allows for general SLOCC operations.
\subsection{Normal form of two-qubit states}
According to Verstraete {\em et al.} it is always possible to obtain a 
Bell-diagonal (renormalised) normal form $\rho^{\mathrm{NF}}$ for two-qubit 
states~\cite{Verstraete2003,Verstraete2001}.
That is, $\rho^{\mathrm{NF}}$ 
can be written as a mixture 
\[
    \rho^{\mathrm{NF}}\ = 
          \ \sum_{j=1}^4 \lambda_j \ket{\phi_j}\!\bra{\phi_j}
\]
where $\lambda_1\geq\lambda_2\geq\lambda_3\geq\lambda_4\geq 0$, $\sum \lambda_j=1$
and $\{\phi_1,\phi_2,\phi_3,\phi_4\}$ is a permutation of the four Bell
states $\{\Phi^+,\Psi^+,\Psi^-,\Phi^-\}$. Evidently the Bell-diagonal
form with maximum $\tilde{\rho}^{\mathrm{NF}}_{00,11}$ is one where 
$\phi_1=\Phi^+$
and $\phi_4=\Phi^-$. It is not difficult to see that by applying appropriate
combinations of the 
local operations $\id_2$, $\sigma_x$, $\sigma_y$, $\sigma_z$ as well as
\[
 \oost
 \left(\begin{array}{cc}
        1   &  1 \\
        1   &  -1 
       \end{array}   
 \right)
\hspace{3mm} \mathrm{and}\hspace{3mm}
 \left(\begin{array}{cc}
        1   &  0 \\
        0   &  i 
       \end{array}   
 \right)
\]
to the qubits in $\rho^{\mathrm{NF}}$, it is always  possible 
to achieve the correct permutation of the $\phi_j$~\cite{Leinaas2006}. 
Note that this implies that there cannot be another Bell-diagonal normal 
form derived from the original two-qubit state $\rho$ with a concurrence
larger than that of $\tilde{\rho}^{\mathrm{NF}}$.

Bennett {\em et al.} have demonstrated that the concurrence of
a Bell-diagonal two-qubit density matrix depends only on its largest 
eigenvalue~\cite{Bennett1996}. Therefore 
$C(\tilde{\rho}^{\mathrm{NF}})=\max(0,2\lambda_1-1)$
can be determined exactly without reference to the Wootters-Uhlmann 
method~\cite{Wootters1998,Uhlmann2000} 
and does not change on applying the symmetrization 
operation equation~\eqref{eq:symmstate} in the main text.

We mention that in the two-qubit case the different optimisation criteria
for step {\em (2)}
({\em i.e.}, maximal concurrence of $\tilde{\rho}^{\mathrm{NF}}$,
maximal fidelity with $\Phi^+$, maximal $\re\tilde{\rho}^{\mathrm{NF}}_{00,11}$,
and minimal Hilbert-Schmidt distance from $\Phi^+$) are equivalent.
\subsection{Entanglement loss in three-qubit symmetrization}
In the main text we have mentioned that for three qubits one may not expect
to find the exact three-tangle for arbitrary mixed states, and that in general
entanglement is lost in the symmetrization.
This statement is illustrated by the mixtures 
\begin{equation}
            \rho_1=p \ket{\mathrm{GHZ}_+}\!\bra{\mathrm{GHZ}_+}+(1-p)\ket{W}\!\bra{W}
\nonumber
\end{equation}
versus
\begin{equation}
\nonumber
\rho_2=p \ket{\mathrm{GHZ}_+}\!\bra{\mathrm{GHZ}_+}+\frac{1-p}{2}\left(\ket{W}\!\bra{W}+
                                                                     \ket{\bar{W}}\!\bra{\bar{W}}
                                                               \right)
\vspace*{2mm}
\end{equation}
where $\ket{W}=\frac{1}{\sqrt{3}}(\ket{001}+\ket{010}+\ket{100})$,
$\ket{\bar{W}}=\sigma_x^{\otimes 3}\ket{W}$.
It is known~\cite{Eltschka2008,Jung2009} that $\rho_1$ has non-vanishing
three-tangle for $p \gtrsim 0.627$,
as opposed to $\rho_2$ which is GHZ-entangled only for $p>3/4$.
Note that $\rho_2$ is already given in the normal form.
For $\rho_1$ the normal form can be calculated analytically.

In the range $0.70<p< 0.74$ the exact three-tangle of $\rho_1(p)$ 
is $0.19<\tau_3(\rho_1(p))<0.31$~\cite{Viehmann2012APB} while $\tau_3(\rho_2(p))=0$.
The optimisation leaves
$\rho_1^{\mathrm{NF}}/\tr \rho_1^{\mathrm{NF}}$ and $\rho_2$ practically 
 unchanged. 
The corresponding points in the 
$xy$ plane are located close to each other and still in the region of $W$
states, {\em i.e.}, we obtain the estimates
   $ \tau_3(\rho^{\mathrm{S}}(\rho_1^{\mathrm{NF}}))=
                     \tau_3(\rho^{\mathrm{S}}(\rho_2))=0$.
Hence, the GHZ entanglement in $\rho_1^{\mathrm{NF}}$ will be 
underrated while that of $\rho_2$ is determined exactly.
\subsection{Relation between projective GHZ witness and quantitative witness}
In the discussion part of the main text we mention that the standard 
projective GHZ witness
$\mathcal{W}_3=\frac{3}{4}\id_8-\ket{\mathrm{GHZ}_+}\!\bra{\mathrm{GHZ}_+}$
can, in modified form, be used as a quantitative witness. Here we explain 
this fact in more detail.

The standard witness $\mathcal{W}_3$ detects the GHZ-type entanglement
in an arbitrary three-qubit state $\rho$: it is a GHZ-class state if
$\tr(\mathcal{W}_3\rho) < 0$. Our aim is to elucidate that 
$\mathcal{W}^{\prime}_3=-4\mathcal{W}_3$ is a quantitative witness for 
$\rho$, {\em i.e.}, that
\begin{equation}
\tau_3(\rho)\ \ge\ \tr(\mathcal{W}^{\prime}_3\rho)\ =\ -4\tr(\mathcal{W}_3\rho) 
\label{SIeq:qwit}
\end{equation}
is a lower bound to $\tau_3(\rho)$ for arbitrary three-qubit states $\rho$.

It appears obvious that, in order to obtain a non-optimal witness, it is not necessary 
to use the GHZ/$W$ line which is difficult to handle analytically. 
The solution of the two-qubit case suggests the following simpler alternative:
We start at the end point of the GHZ/$W$ line $P=(x=\frac{3}{8},y=\frac{\sqrt{3}}{6},\tau_3=0)$ 
and consider the straight line which contains this point and is parallel
to the lower-left border of the triangle. Its equation is 
$y_P=\left( -2x+\frac{5}{4}\right)/\sqrt{3}$. It crosses  the triangle only
in the GHZ part, that is, its points lie above the GHZ/$W$ line. For all the states 
$\rho^{\mathrm{S}}_P$ which correspond to triangle points
on this line the Hilbert-Schmidt scalar product with GHZ$_+$ 
equals 
$(\mathrm{GHZ}_+,\rho^{\mathrm{S}}_P)\equiv
      \frac{1}{2}\tr\left(\ket{\mathrm{GHZ}_+}\!\bra{\mathrm{GHZ}_+} \rho^{\mathrm{S}}_P
                    \right)
                                             =\frac{3}{8}$.

From Supplementary Fig.~S2 it is easy to see that a plane which 
contains this line and the point 
$(x_{\mathrm{GHZ}_+}=\frac{1}{2},y_{\mathrm{GHZ}_+}=\frac{\sqrt{3}}{4},\tau_3=1)$
represents a lower bound to the three-tangle of GHZ-symmetric three-qubit states.
It is straightforward to check that the function 
$\rho^{\mathrm{S}}\mapsto \tau_3^P(\rho^{\mathrm{S}})$ corresponding
to the points of that plane is given by
\begin{equation}
     \tau^P_3(\rho^{\mathrm{S}})\ =
  \  -4\tr\left(\mathcal{W}_3\rho^{\mathrm{S}}\right)\
      \ \ .
\label{SIeq:standardwitness}
\end{equation}
As the plane $\tau_3^P(\rho^{\mathrm{S}})$ lies below the exact $\tau_3(\rho^{\mathrm{S}})$ 
we have also
\[
     \tau^P_3(\rho^{\mathrm{S}})\ \le
      \ \tau_3(\rho^{\mathrm{S}})\ \ .
\]
Further, the operator $\mathcal{W}_3$ has GHZ symmetry so that for an arbitrary state $\rho$
\[
   \tr\left(\mathcal{W}_3\rho\right)\ =\
   \tr\left(\mathcal{W}_3\rho^{\mathrm{S}}(\rho)\right)\ \ .
\]
By combining the preceding relations and the conclusions from the Section 
``Results'' in the main text we obtain
\begin{align}
-4\tr\left(\mathcal{W}_3\rho\right)  =
-4\tr\left(\mathcal{W}_3\rho^{\mathrm{S}}(\rho)\right)  
    & \le  \tau_3\left(\rho^{\mathrm{S}}(\rho)\right)
\nonumber\\
    & \le  \tau_3\left(\rho^{\mathrm{S}}(\tilde{\rho}^{\mathrm{NF}})\right)
\nonumber\\
    & \le  \tau_3(\rho)
\label{SIeq:standardwitness2}
\end{align}
which confirms the desired result, equation~\eqref{SIeq:qwit}. We mention that also
here there is a certain freedom whether or not one wants to optimise the state
$\rho$ before symmetrizing it. 
One may note the relation between this type of equation deriving from
our method and some of the
findings in Sections 3.4--3.6 of Ref.~\cite{Eisert2007}, 
as well as those in Ref.~\cite{LS2012}
\begin{figure}[bht]
  \centering
  \includegraphics[width=.70\linewidth]{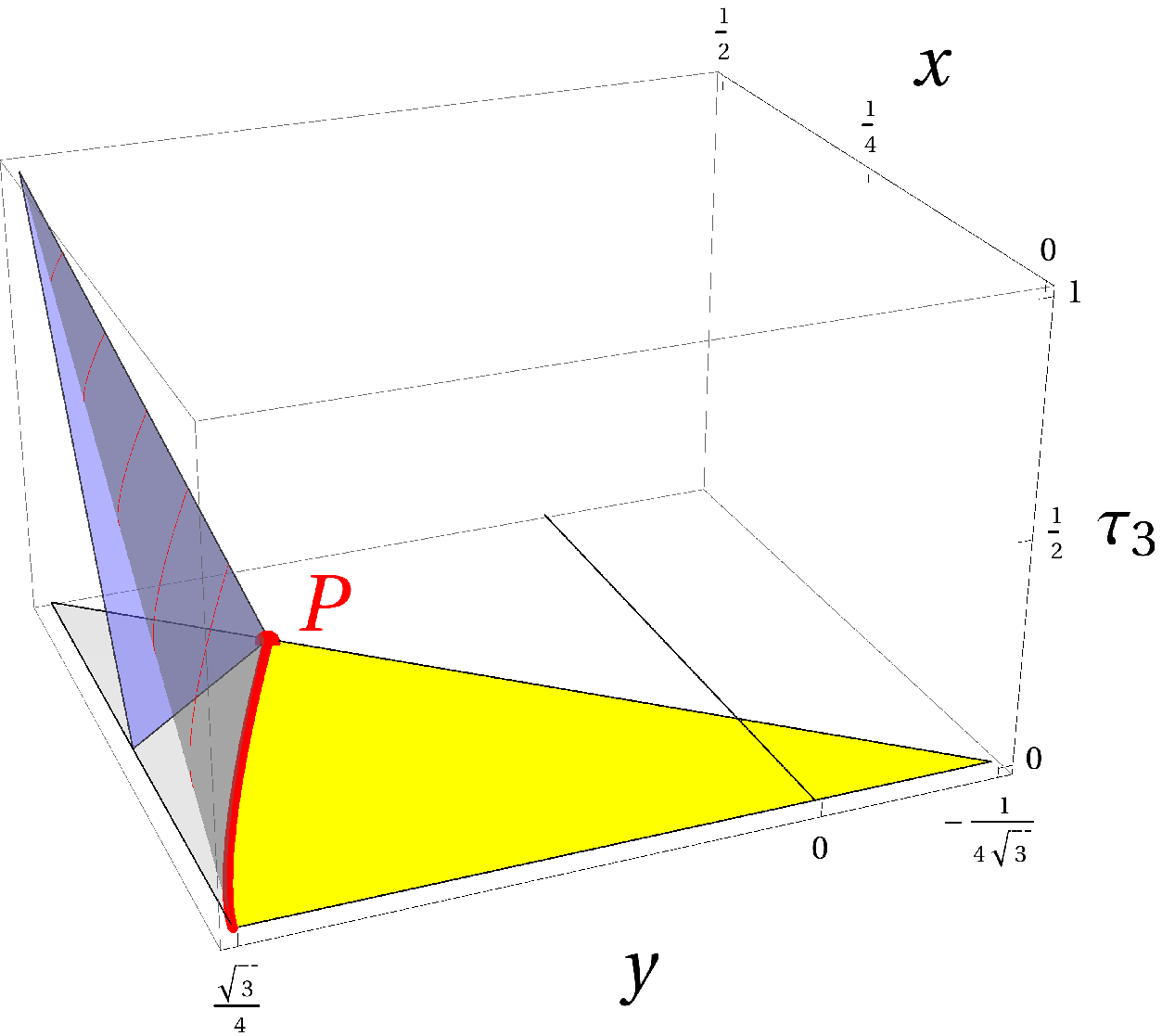}
  \caption{
           The three-tangle for three-qubit GHZ-symmetric states
           as in 
           equation~\eqref{eq:result3} (grey surface)
           compared to the non-optimal (but easy-to-handle) 
           quantitative witness, supplementary 
           equation~\eqref{SIeq:standardwitness}, blue triangle (see text).
            } 
\end{figure}

From these remarks one might feel tempted to conclude that our method
is a mere extension to the standard witness approach as it detects GHZ
entanglement in a given state $\rho$ more or less according to the 
fidelity of the GHZ state and assigns a number to it. To clarify this point
consider the example
\[
     \rho_3\ =\ 
            p \ket{\mathrm{GHZ}_+}\!\bra{\mathrm{GHZ}_+}+(1-p)\ket{001}\!\bra{001}
            \ \ .
\]
The exact three-tangle is $\tau_3(\rho_3)=p$, that is, the state contains GHZ
entanglement for arbitrarily small $p$. While the standard witness would not
detect entanglement for $p<3/4$ our approach produces the correct value
(with a relative error $< 10^{-2}$) for values as small as $p\sim 10^{-5}$.
\end{document}